\documentclass[10pt, conference, compsocconf]{IEEEtran}

\usepackage[utf8]{inputenc} % allow utf-8 input
\usepackage[T1]{fontenc}    % use 8-bit T1 fonts
\usepackage{url}            % simple URL typesetting
\usepackage{booktabs}       % professional-quality tables
\usepackage{amsfonts}       % blackboard math symbols
\usepackage{nicefrac}       % compact symbols for 1/2, etc.
\usepackage{microtype}      % microtypography
\usepackage{lipsum}
\usepackage{graphicx}
\usepackage{amsmath}
\usepackage{tikz}
\usepackage{listings}
\usepackage{amsmath}
\usetikzlibrary{automata}
\graphicspath{ {./images/} }

\usepackage{algorithm}% http://ctan.org/pkg/algorithm
\usepackage{algpseudocode}% http://ctan.org/pkg/algorithmicx

\usepackage{todonotes}
\usepackage{longtable}
\makeatletter
\def\BState{\State\hskip-\ALG@thistlm}
\makeatother

\definecolor{codegreen}{rgb}{0,0.6,0}
\definecolor{codegray}{rgb}{0.5,0.5,0.5}
\definecolor{codepurple}{rgb}{0.58,0,0.82}
\definecolor{backcolour}{rgb}{0.95,0.95,0.92}

\begin{document}

\title{CodeDSI: Differentiable Code Search}

\author{
\IEEEauthorblockN{Usama Nadeem} 
\IEEEauthorblockA{
School of Information Technology \\
Illinois State University\\
unadee2@ilstu.edu
}
\and
\IEEEauthorblockN{Noah Ziems} 
\IEEEauthorblockA{
School of Information Technology \\
Illinois State University\\
nziems@ilstu.edu
}
\and
\IEEEauthorblockN{Shaoen Wu}
\IEEEauthorblockA{
Department of Information Technology \\
Kennesaw State University \\
swu10@kennesaw.edu}

}
% =============== TODO LIST ===============
% 

\maketitle
\begin{abstract}
Reimplementing solutions to previously solved software engineering problems is not only inefficient but also introduces inadequate and error-prone code. Many existing methods achieve impressive performance on this issue by using autoregressive text-generation models trained on code. However, these methods are not without their flaws. The generated code from these models can be buggy, lack documentation, and introduce vulnerabilities that may go unnoticed by developers. An alternative to code generation—neural code search— is a field of machine learning where a model takes natural language queries as input and, in turn, relevant code samples from a database are returned. Due to the nature of this pre-existing database, code samples can be documented, tested, licensed, and checked for vulnerabilities before being used by developers in production. In this work, we present CodeDSI, an end-to-end unified approach to code search. CodeDSI is trained to directly map natural language queries to their respective code samples, which can be retrieved later. In an effort to improve the performance of code search, we have investigated docid representation strategies, impact of tokenization on docid structure, and dataset sizes on overall code search performance. Our results demonstrate CodeDSI strong performance, exceeding conventional robust baselines by 2-6\% across varying dataset sizes.
\end{abstract}

\section{Introduction}\label{sec:introduction}
Reimplementing solutions to previously solved problems is not only inefficient but also introduces error-prone code. Previous approaches to this problem achieve impressive performance using autoregressive text-generation models trained on code \cite{codex}. However, these methods are not without their flaws. The generated code from these models can be buggy, untested, lack documentation, and introduce vulnerabilities that may go unnoticed by developers. An alternative to code generation—neural code search— is a field of machine learning where a model takes natural language queries as input and, in turn, relevant code samples from a database are returned. Due to the nature of this pre-existing database, code samples can be documented, tested, licensed, and checked for vulnerabilities before being used by developers in production.

In code search, a natural language query $q \in Q$ is mapped to a ranked list of relevant code samples $\{c_1, \dots, c_n\} \subseteq C$. This is similar to many information retrieval(IR) systems which map natural language queries to natural language documents $\{d_1, \dots, d_n\} \subseteq D$. However, although words in $q$ can often be found in many relevant documents $d \in \{d_1, \dots, d_n\}$ for IR, words in $q$ are rarely found in relevant code samples $c \in \{c_1, \dots, c_n\}$. The added complexity from the query-code distribution mismatch means simple word-frequency based solutions such as TF-IDF \cite{tfidf} often work well for traditional IR, but rarely work well for code search.

State-of-the-art approaches to code search solve the distribution mismatch problem by relying on dual encoders(DEs) \cite{codebert,graphcodebert,scalable sentence encoders,end-to-end retrieval,neural ranking} to measure distance between a query $q$ and code sample $c_i$ for all $c \in C$, then apply a nearest neighbor search to find the $n$ most relevant code samples $\{c_1, \dots, c_n\} \subseteq C$. However, recent work in information retrieval (IR) has shown significant performance improvement upon DEs by using a sequence-to-sequence(seq2seq) architecture \cite{attention}, known as \textit{differentiable search index}(DSI) \cite{dsi}, to directly map natural language queries $q \in Q$ to a relevant document identifier(docid) $j \in Y$. With DSI, the content of each $d \in D$ is encoded within the weights of a seq2seq architecture. During inference, the model is trained to map a natural language query $q$ to a docid $j$.

In this work, we propose a DSI model, CodeDSI, for code search and show that DSI significantly outperforms DEs in code search while also being much better at handling the code-query distribution mismatch problem. Our main contributions are as follows:

\begin{enumerate}
    \item First, we develop CodeDSI and show that CodeDSI significantly improves upon previous code search SOTA on varying datasets.
    
    \item Second, we investigate that docid representation plays a crucial role in code search, with clustering outperforming other naive approaches.
    
    \item Third, we examine that semantic clustering further improves performance and may provide intuitive document characterization.
    
    \item Forth, we analyze the influence of tokenization on docid structure as it relates to overall code search performance.
\end{enumerate}

\section{Related Work} \label{sec:relatedwork}
The works that relate to ours can largely be split up into two main
categories: Information retrieval and NLP on artificial languages.

\subsection{Information Retrieval}
Beginning with Differentiable Search Index (Tay et al., 2022) \cite{dsi}, a text-to-text model that maps string queries directly to relevant docids. Our research is an extension of DSI, which originally presented the capability of storing indexes in model parameters, and then retrieving the appropriate docids during inference. The DSI paper only tested on a Natural Questions dataset(Kwiatkowski et al., 2019) \cite{natural questions}, which consisted of question-answer pairs. We take this one step forward, using DSI on code search, where we believe this approach can be best exercised. Additionally, we not only experiment with docid representation strategies, but also docid structure and the influence of tokenization on such structure.   

To note, there has been prior work on learned indexes (Roberts et al., 2020) \cite{memory store} that used large Transformer models as a memory store. However, the approach presented here targets to generate direct answers to queries, whereas we aim to generate docids to retrieve the appropriate documents.

There is also substantial work on retrieval augmented generation (Borgeaud et al., 2021; Guu et al., 2020) \cite{improving language by tokens,REALM}, where auxiliary documents are retrieved to enhance generation. Here, the task of retrieval is used to augment the generation process. For example, during execution, the model retrieves documents from an external dataset in conjunction with the provided query to generate an output. 

Alternatively, text-generation can be used to directly enhance the provided query, and in turn, improve retrieval task (Nogueira et al. 2019) \cite{document expansion}. For example, before indexing, the document is passed to a seq2seq model which generates additional text, such as questions pertaining to the document, that can be useful in retrieval.

Furthermore, Dual Encoders are also well known for their expertise in retrieval (Feng et al., 2020; Guo et al., 2021; Ni et al., 2021; Gillick et al, 2018. Dehghani et al, 2021) \cite{codebert,graphcodebert,scalable sentence encoders,end-to-end retrieval,neural ranking}. The idea behind Dual Encoders is rather simple and effective: Generate query and document embedding independently using a sequence encoder, and then perform a similarity comparison to retrieve the appropriate document. This approach is different from our goal as we intend for the model to memorize the information about each document, map it to its appropriate docid during pretraining, and finally, retrieve the appropriate docid at inference time.

Finally, De Cao et al. (2020) \cite{autoregressive} presented a related seq2seq architectural design, autoregressive entity linking. Here documents mentioning an entity are mapped to a canonical name of that entity. Therefore, enabling the retrieval of appropriate documents based on the canonical name(title) provided. This approach is different from our goal as we try to retrieve documents containing the relevant code sample, not whose title may be an answer. Furthermore, autoregressive entity linking aims to provide a semantic identifier, whereas our approach allows such identifiers to be arbitrary.

\subsection{NLP on Artifical Languages}
% Papers that use NLP on Code(CodeBERT, CodeT5, etc.....)
CodeBERT (Feng et al., 2020) \cite{codebert} is a bimodal pretrained model that apprehends the semantic link between natural language and programming language, producing general-purpose representations to support NL-PL applications. CodeBERT is a multilayer transformer based architecture and is trained with a hybrid objective function that incorporates the pretraining task of replaced token detection, enabling it to learn NL-PL cross-modal representation. CodeBERT is a pioneer model, and we will be using it as a baseline to compare with our results.

CodeT5 (Wang et al., 2021) \cite{codet5}, an extension of T5, is a seq2seq architecture that supports code understanding, generation tasks, and multi-task learning. CodeT5 is a unified pretrained encoder-decoder transformer model that is pretrained with a novel identifier-aware task that enables the model to determine which code tokens are identifiers and to recover them when masked. CodeT5 supports many programming tasks, such as text-to-code generation, code autocompletion, code summarization, code translation, defect detection, and clone detection. However, CodeT5 does not extend to code search, which is due to its limitation as a seq2seq architecture design. In this paper, we extend CodeT5 to support code search using Differentiable Code Search. Given CodeT5 success in many programming applications, we believe CodeT5 will outperform prior methods on code search as well.

Some other approaches, to better the performance of code search, employ improving the vector space by using tree-serialized representations and multimodal learning models. Gu et al. (2022) \cite{multimodal representation}, introduce a tree-serialization method on a simplified form of AST, and build a multimodal representation for the code data. TransCoder (Lachaux et al., 2020) \cite{transcoder} explores unsupervised programming language translation, showcasing the model understanding of code across various languages. Finally, more recently, GraphCodeBERT (Guo et al., 2021) \cite{graphcodebert} introduced a pretrained model for programming languages that considers the inherent structure of code, providing crucial code semantics and enhanced code understanding.

\begin{figure*}[h]
  \includegraphics[scale=.6]{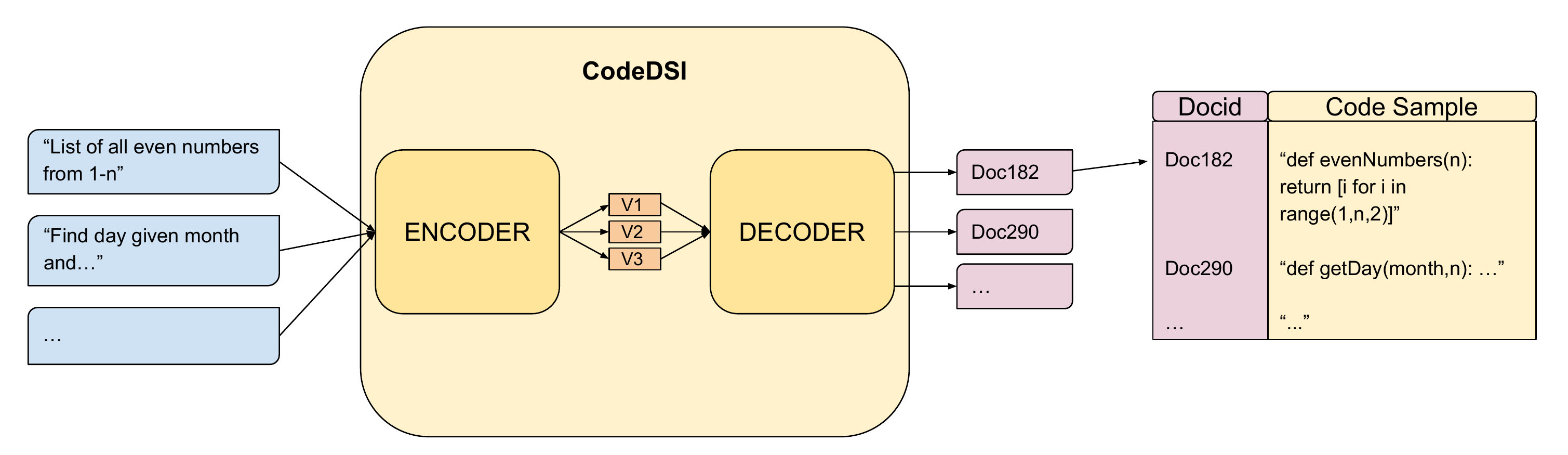}
  Figure 1: Above is a flow diagram displaying the unified encoding and retrieval task performed by CodeDSI. Here, during inference, the queries are given as input and the docids are directly produced by the seq-to-seq model. Each docid corresponds to their respective code sample as indicated by the sample database on the right. The assigned docids can be arbitrary or semantic in nature.
\end{figure*}

\section{Differentiable Code Search} \label{sec:dataset}
As in traditional IR systems, Differentiable Code Search can be broken down into two separate steps. Namely, these are \textit{Indexing} and \textit{Retrieval}. Where traditional IR systems, including DEs, do these in two separate parts, Differentiable Code Search does this all in a single forward pass during evaluation. Differentiable Code Search breaks these steps into two tasks which can then be trained jointly or separately.

\subsection{Indexing}
The task of indexing is training the model to memorize, or index, the information about each document and map to its appropriate docid. Here the index is stored in the model parameters, and indexing becomes a part of model training, enabling the model to learn which documents belong with which identifiers, and effectively retrieve them later. To train the model to do Indexing, the model is given a code sample or part of a code sample as input, and is trained to predict the docid of the code sample in string format.

\subsection{Retrieval}
Similar to how the model is trained for indexing, to train the model for Retrieval, a query is given as input and the model predicts the docid of the code sample in string format.

By first training the model to Index, the model learns the mapping of code samples to their respective docids. When the model is later trained for Retrieval, the contents of the code samples and their mappings are already known. In other words, the content of the code samples are memorized by the model.

\subsection{DocID Representation}
To enable the model to effectively establish the association between query-code pairs, we structure docid using several strategies: Considering both semantic and naive approaches to docid representation.  

\begin{enumerate}
    \item \textbf{Direct Representation}\\
    A naive approach to generating docids. Create \textit{X} arbitrary unique identifiers to represent the respective code samples, with sequential order preserved.

    \item \textbf{Clustered Representation}\\
    A semantic approach to generating docids. Applying K-Means clustering over embedding generated by CodeBERT. Creating 10 clusters, each document is recursively assigned a unique value from 0-9 to ultimately generate a semantically unique docid.
\end{enumerate}

The naive approach to docid representation, Direct Representation, serves as a baseline, given its simplicity and forthright approach to docid representation. The semantic approach to docid representation, Clustered Representation, explores the idea of extracting knowledge from the problem domain. Here by using CodeBERT, we obtain vector embeddings of each code sample. Then, using K-Means clustering, we can achieve like-clusters, where embeddings closer in space are grouped together. Therefore, ultimately, code samples that provide solutions to similar queries are given comparable identifiers. 

Furthermore, we also explore structuring such document identifiers numerically and alphabetically, to exercise the influence of tokenization. Our experiments indicate that numerical identifiers tend to perform better than alphabetical identifiers. We believe this may be due to the nature of tokenizers, where numerical identifiers are provided a unique token per digit. However, strings of characters may simply be represented under a single token, therefore creating a lack in strong independent identifiers.      

\begin{algorithm}
\caption{Clustered Representation}\label{euclid}
\hspace*{\algorithmicindent} \textbf{Input:} \texttt{Document Embedding X, Empty array DocIDs} \\
\hspace*{\algorithmicindent} \textbf{Output:} \texttt{Corresponding DocID array DocIDs} \\
\begin{algorithmic}[1]
\Function{cluster}{$X,DocIDs$}
    \If{$len(X) < 100$}
        \State $DocIDs \gets generateNaiveIDs(X)$
        \State \Return $DocIDs$
    \EndIf \\
    \State $Y \gets KMeansCluster(X, k=10)$
        \If{$DocIDs$ is empty list} 
            \State $DocIDs \gets Y$
        \Else
            \For{\texttt{i,value in Y}}
                \State $DocIDs[i] \gets value$
              \EndFor
        \EndIf \\
        
        \For{\texttt{i=0 to 9}}
            \State $Subcluster_i \gets Y==i$
            \State $DocIDs[Subcluster_i] \gets
            \Call{cluster}{X[Subcluster_i],DocIDs[Subcluster_i]}$
          \EndFor \\
          \State \Return $DocIDs$
\EndFunction
\end{algorithmic}
\end{algorithm}

\section{Experiments}\label{sec:model}
\subsection{Datasets}\label{subsec:datasets}

%This is what I added/reworded:
For our testing purposes, we use the CodeXGlue Natural Query Search dataset (Shuai Lu et al., 2021) \cite{CodeXGlue}, which consists of X query-code pairs. This particular dataset consists of only Python code, but no preprocessing specific to Python is done. For further experimentation, and official results, we use the CodeSearchNet Corpus dataset (Hamel Husain et al., 2019) \cite{CodeSearchNet}, which is a dataset of 2 million query-code pairs from open-source libraries. It contains code samples spanning six programming languages (Go, Java, JavaScript, PHP, Python, and Ruby). Raw code samples do include documentation, but we remove such documentation during our preprocessing to eliminate any bias.

\subsubsection{CodeXGlue}
% Describe CodeXGlue here
CodeXGlue WebQueryTest is a dataset consisting of real code search queries for Python code samples. Data is collected through the web query logs of a search engine. Collected user queries are filtered by the keyword ‘python’ and excludes queries without code search intent. Data is then shrunk by selecting the top candidates with the highest query-code similarity computed by a CodeBERT-base. The final dataset has been annotated with whether the collected code (with its documentation) can answer the respective query.

\subsubsection{CodeSearchNet}
% Describe CodeSearchNet here
CodeSearchNet Corpus is a dataset of 2 million query-code pairs. The data collected is from publicly available open-source non-fork GitHub repositories: This excludes code samples that are not licensed or whose license does not permit redistribution. The collected code samples vary across six programming languages (Go, Java, JavaScript, PHP, Python, and Ruby). The dataset has been preprocessed where documentation is truncated to make the length more comparable, removal of special methods (constructors, extension methods, etc), and elimination of near-duplicate functions.

\subsection{Statistical Baseline}
Term frequency-inverse document frequency, TF-IDF, is a mathematical statistic used to determine the importance of a word to a document in a collection \cite{tfidf}. This is a non-parametric algorithm for search and will be used as another baseline for our comparison. TF-IDF, as defined in Eq. (\ref{eq:tfidf}), has 2 parts: Term frequency as defined in Eq. (\ref{eq:tf}), and Inverse document frequency Eq. (\ref{eq:idf}). Term frequency, as the name suggests, calculates the weight of a document by the number of relevant terms that occurs in such a document. 

\begin{equation} \label{eq:tfidf}
    \textbf{TF-IDF}=TF(term,doc)*IDF(term) 
\end{equation}

\begin{equation} \label{eq:tf}
    \textbf{TF}=\frac{FreqOfTerm \in doc}{No.OfWords \in doc} 
\end{equation}

\begin{equation} \label{eq:idf}
    \textbf{IDF}=log(1+\frac{No.OfDocs}{No.OfDocsWithTerm})
\end{equation}

For example, when searching for a code sample most relevant to the query, ``List of all even numbers from 1 to \textit{n}", the algorithm would first break the query into its individual terms, and then search documents relevant to these keywords. To further distinguish these documents, the algorithm also accounts for the frequency of each term in the document. Inverse document frequency refers to the quantified specificity of a term in the documents in which it occurs. Recognizing that some terms are more common than others, and therefore are poor keywords to identify relevant documents. Hence, such terms are given less weight when determining the best document to query. Together, TF*IDF, determines the relative frequency of terms in a document contrasted with the inverse proportion of that term over all documents in a collection.

For our code, we first create a frequency map of the raw code samples by running it through the document collection. From this, we convert our mapping into a bag-of-word representation. Then using each input query, we pass it to TF-IDF, as described above, to generate a vectorized representation of each document. Finally, we can extract the most relevant document predicted by TF-IDF and compare it to the given label.

\subsection{Implementation details}
All CodeDSI models are initialized with the same default hyperparameters and model configuration. Therefore, all tests include a CodeT5-base-multi-sum as our foundation and size. For our choice of tokenizer, we use CodeT5's code-specific tokenizer provided with our selected model size. For our choice of dataset, we use CodeSearchNet for its large collection and wide variety of languages. To note, we choose to use all 6 available languages during our testing for all experiments. Due to the limited hardware and resources available, we experiments on only a portion of the complete dataset obtained. In regards to the semantic docid representation, we use CodeBERT-base in correspondence with Sckit-Learn K-Means clustering to generate the semantic docids. We present results for all docid representation strategies and structure.

\section{Results and Analysis}\label{sec:performance}

\begin{table*}[]
\centering
\resizebox{15cm}{!}{
\begin{tabular}{|lll|lll|}
\hline
\multicolumn{3}{|c|}{Model Parameters}                                              & \multicolumn{3}{c|}{CodeSearchNet}                                                             \\ \hline
\multicolumn{1}{|l|}{Model}    & \multicolumn{1}{l|}{Size}           & Method       & \multicolumn{1}{l|}{Dataset1k}        & \multicolumn{1}{l|}{Dataset10K}       & Dataset50K       \\ \hline
\multicolumn{1}{|l|}{TF-IDF}   & \multicolumn{1}{l|}{-}              & -            & \multicolumn{1}{l|}{5.87\%}           & \multicolumn{1}{l|}{4.71\%}           & 3.00\%           \\
\multicolumn{1}{|l|}{CodeBERT} & \multicolumn{1}{l|}{Base}           & Dual Encoder & \multicolumn{1}{l|}{92.14\%}          & \multicolumn{1}{l|}{89.80\%}          & 66.84\%          \\
\multicolumn{1}{|l|}{CodeDSI}  & \multicolumn{1}{l|}{Base-Multi-Sum} & Naive-int    & \multicolumn{1}{l|}{\textbf{94.27\%}} & \multicolumn{1}{l|}{89.28\%}          & \textbf{73.26\%} \\
\multicolumn{1}{|l|}{CodeDSI}  & \multicolumn{1}{l|}{Base-Multi-Sum} & Naive-char   & \multicolumn{1}{l|}{93.65\%}          & \multicolumn{1}{l|}{86.97\%}          & 60.58\%          \\
\multicolumn{1}{|l|}{CodeDSI}  & \multicolumn{1}{l|}{Base-Multi-Sum} & Cluster-int  & \multicolumn{1}{l|}{\textbf{94.79\%}} & \multicolumn{1}{l|}{\textbf{90.44\%}} & \textbf{72.44\%} \\
\multicolumn{1}{|l|}{CodeDSI}  & \multicolumn{1}{l|}{Base-Multi-Sum} & Cluster-char & \multicolumn{1}{l|}{93.85\%}          & \multicolumn{1}{l|}{84.32\%}          & 71.37\%          \\ \hline
\toprule

\multicolumn{6}{l}{%
  \begin{minipage}{11cm}% 
    Table 1: Experimental results on CodeSearchNet. Baseline includes TF-IDF and Dual Encoder, which are outperformed by CodeDSI. We train CodeDSI with various documentation strategies, with naive-int and cluster-int performing best.
  \end{minipage}%
}\\
\end{tabular}
}
\end{table*}

\subsection{Experimental Results}
Table 1 reports results on CodeSearchNet for 1K, 10K, and 50K dataset sizes. The reported results are an absolute measure of accuracy of the model predictions. The \textit{size} and \textit{method} columns indicate the absolute model size and approach used for CodeSearchNet evaluation. Here, CodeBERT uses the traditional DE method and CodeDSI predicts document label using our various docid representation strategies. 

\subsection{Baseline Results}
TF-IDF and CodeBERT are our baseline for their tradition approaches to code search. TF-IDF uses a statistical approach to determining the relative query-code pairs. As expected, TF-IDF performs poorly due to its strong dependence on relevant keywords provided by the given query. However, because we remove all documentation from the collected code samples, TF-IDF can only depend on thoughtful variable and function names to determine query-code pairs. On the other hand, CodeBERT produces satisfactory results. CodeBERT uses the conventional dual encoder method to determine query-code pairs: Mapping code and query samples as vectors in a common space, producing vector encodings. Then finding the most relevant label during the retrieval stage using a maximal inner product search.

\subsection{CodeDSI Results}
Our results for CodeDSI outperform our baselines, TF-IDF and dual encoder, across all dataset sizes. As hypothesized, the performance of CodeDSI fluctuates as the size of the dataset is increased: With CodeDSI performing best on the smaller dataset sizes 1K and 10K, and that gap widening as we approach 50K.   

\subsubsection{Docid Representation \& Structure}
One key concern in this paper is in regards to document identifier representation. We experimented with four variants of docid representations strategies: Naive-int, Naive-char, Cluster-int, and Cluster-char. The Naive-int method is a direct representation approach with a numerical docid structure. This approach performed consistently across all dataset sizes. Naive-char, another direct representation approach, underperformed in comparison to its counterpart. We believe this substantial gap between the two similar representation strategies is due to the influence of tokenization. Tokenizers often represent numbers with distinct tokens for each digit. However, many characters together create a string that can be represented under a single token, therefore, creating a lack in strong independent document identifiers. Similarly Cluster-char, a clustered representation approach, underperformed in comparison to Cluster-int. Further reinforcing the impact on how docid representation and structure can greatly influence model performance.

\subsubsection{Direct vs Clustered Approach}
In this section, we explore the performance of Direct vs Clustered approach to docid representation. While direct offers no semantic docid structure, it does performs well across all dataset sizes. Clustered, similarly, also performs strongly across all dataset size, outperforming direct in many cases. We believe the semantic structure to docid representation produced by the clustered approach creates more robust docid representation, given the stability in performance as the datasize increases: Particularly notable at 50K, where the Naive-char method achieves a low 60.58\%, whereas Cluster-char obtains a 71.37\%. We hypothesize that the clustered approach may also be more resilient to catastrophic forgetting. However, we defer this line of investigation for future work.

% Talk about clustering here

\section{Conclusion}\label{sec:conclusion}
In this paper we propose Differentiable Code Search, CodeDSI, an end-to-end unified approach to code search. We present various docid representation strategies expanding naive and semantic approaches. We survey the impact of tokenization on docid structure, exploring both numeric and character representation for such document identifiers: Showcasing that numeric document identifier outperform their counterpart. In regards to direct and clustered approach to docid representation, we note consistent performance, with semantic docids outperforming naive approaches. We also analyze the impact of dataset sizes on model performance. As the datasize is increased, the model becomes more prone in its absolute performance. 

Although, we believe our model and results are very promising, we still have further work that needs exploring. For example, though the semantic and naive approach perform similarly, we hypothesize semantic document identifier representation may be more resilient to catastrophic forgetting. We also believe the interdependence of tokenization and docid structure can be further surveyed in regards to overall model performance. Furthermore, we believe an even greater variety of document structure can be investigated, with a mix of both numeric and character combination to create further robust document identifiers. Finally, it may also be interesting to analyze CodeDSI performance in regards to each specific programming language.

\section{Acknowledgment}\label{sec:acknowledgment}
This material is based upon work supported by the National Science Foundation under Grant No.2109971.

\end{document}